\documentclass[11pt]{article}                                                 
\usepackage{amsfonts}             
\usepackage{amsmath,amssymb,epsfig}

\font\bbold=bbm10
\def\mC{\hbox{\bbold C}}
\def\mP{\hbox{\bbold P}}
\def\mCP{\hbox{\bbold CP}}

\def\CA{{\cal A}}
\def\CZ{{\cal Z}}
\def\CN{{\cal N}}
\def\CM{{\cal M}}
\def\CP{{{\mathbb C}{\mathbb P}}}

\def\eg{{\it e.g. }}
\def\ie{{\it i.e. }}

\begin{document}

\begin{flushright}
LMU-ASC 76/05
\end{flushright}

\begin{center}
\vspace{2cm}
{ \LARGE {\bf Twistor spaces, mirror symmetry and \\ \vspace{.2cm}
    self-dual K\"ahler manifolds}} 

\vspace{1.5cm}

Giuseppe Policastro

\vspace{0.8cm}

{\it Arnold Sommerfeld Center, Department f\"ur Physik \\
Ludwig-Maximilians-Universit\"at M\"unchen \\
Theresienstra{\ss}e 37, D-80333 M\"unchen, Germany
}

\vspace{0.3cm}
policast@theorie.physik.uni-muenchen.de
\vspace{2cm}

\end{center}

\begin{abstract}

We present the evidence for two conjectures related to the twistor string. The first conjecture states 
that two super-Calabi Yaus -- the supertwistor space and the superambitwistor space -- form a mirror pair. 
The second conjecture is that the B-model on the twistor space can be
seen as describing a 4-dimensional gravitational theory, whose
partition function should involve a sum over ``space-time foams''
related to D1 branes in the topological string.  

This article is based on a talk given at the International Workshop ``Supersymmetries and 
Quantum Symmetries'', 27-31 July 2005, JINR, Dubna, Russia.

\end{abstract}

\vspace{1cm}

\section{Introduction}

Witten's reformulation of  perturbative ${\cal N}=4$ supersymmetric $U(N)$
Yang-Mills theory as a topological string theory
with the supertwistor space as target \cite{Witten} has inspired a large 
number of works that have extended the original insight and fulfilled one of the initial expectations, namely 
that of simplifying the calculations of YM amplitudes by exploiting the holomorphic properties they 
exhibit in twistor space (see \cite{Cachazo} for a review of these
results and a fuller list of references). This achievement was in
keeping with the general philosophy of the twistor program, which has
at its starting point the observation that complicated field equations
in Minkowski space can be recast in the form of simpler cohomology
problems via the so-called Penrose transform (\cite{Penrose} and
\cite{WardVS} for a general presentation). 

By contrast, some puzzles 
that were pointed out by Witten from the beginning are still unsolved. Probably the most important open question  is that of the coupling (or decoupling) of the closed string modes. While they can be ignored at tree level, they inevitably contribute in loops and one has to find a way of disentangling them from field theory, or else of making sense of field theory coupled to conformal supergravity. In other directions there has been some progress, \eg in extending the correspondence to situations with lower supersymmetry, and in setting up the correspondence using the ambitwistor construction, which avoids breaking the parity symmetry between self-dual and anti-self-dual solutions. 

One remarkable aspect of Witten's proposal is the fact that the complete 4-dim gauge theory is described in terms of a topological string; in their usual applications to Calabi-Yau compactifications, topological strings can describe only the F-terms of the corresponding field theory. However the Calabi-Yau appearing
 here as the twistor space is  not part of some internal geometry, but it contains the space-time as well. Moreover, it is a supermanifold. So even though formally the twistor topological string is not different from its more usual incarnations, it encodes the physical information in a rather different way. This begs the question of how much of the common lore concerning these models can be directly applied to this situation.  The most striking result of the investigations of topological strings is the existence of mirror symmetry: Calabi-Yau spaces come in pairs, and two spaces in a mirror pair must satisfy some conditions, \eg relations between their Hodge numbers. It is not entirely obvious whether this should extend to the case of super-CY, and how. One problem is that a complete cohomology theory for supermanifolds is still lacking (for discussions on these issues see \cite{Sethi,schwarz,Grassi}). Nevertheless, this question can be addressed at a formal level, applying the usual techniques of gauged linear sigma model.  In section \ref{sec:mirr} we discuss these issues and present results obtained with S.P. Kumar \cite{Kumar:2004dj}. 
 
In section \ref{sec:foam} we look at the closed string sector, and try to interpret the D1 branes as gravitational sources. This work was done with S. Hartnoll \cite{Hartnoll:2004rv}. It can be argued that the backreaction of the branes on the geometry is related, via the twistor correspondence, to a transition between different gravitational instantons in 4 dimensions. This raises the possibility of having a sector of 4-dim gravity that can be described by a gas of D1-branes; in spirit this would be similar to the quantum foam description of the gravity sector of the A-model topological string on CY 3-folds \cite{Iqbal:2003ds}. Whether such a description can be useful in this case and whether may lead to some exact results remains to be seen.

 \section{Mirror symmetry}\label{sec:mirr}
 
Let us now briefly review the twistor and ambitwistor  constructions and the  arguments in favor of the existence of mirror symmetry put forward in \cite{Neitzke,AganagicYH} and then in \cite{Kumar:2004dj}. 

The (bosonic) twistor space $\CZ={\mCP}^{3}$ is a ${\mCP}^{1}$ fibration over $S^{4}$. The fibre over a point $p$ parametrizes almost-complex structures on $T_{p}S^{4}$. There is a 1-1 correspondence (Penrose-Ward) between rank $r$ holomorphic vector bundles on $\CZ$, trivial when restricted to the fibres, and $SU(r)$ anti-self dual connections on $S^{4}$ (or suitable open sets on both sides). This is, in essence, because an ASD gauge field has a field strength of type $(0,2)$ for any choice of complex structure compatible with the orientation. Topologically, the bundle on $\CZ$ is the pull-back of the gauge bundle on $S^{4}$, and the holomorphic structure is given by a $(0,1)$ connection $\CA$ satisfying $\bar \partial \CA + \CA \wedge \CA =0$. 
The latter are the field content and the equations of motion of holomorphic Chern-Simons theory, which in turn is the effective space-time theory of the open string sector of the topological B-model on a Calabi-Yau. The action is  

\begin{equation}\label{hCS}
S = \int_{\CZ} \Omega \wedge (\CA \wedge \bar \partial \CA + \frac23 \CA \wedge\CA\wedge\CA) \,. 
\end{equation}

Here $\Omega$ is a $(3,0)$ form. $\CZ$ is not a Calabi-Yau, but the problem is circumvented by adding fermionic coordinates. The total space is then taken to be ${\mCP^{3|4}}$, which can also be seen as a fermionic bundle $\Pi({\cal O}(1)\otimes {\mathbb C}^{4})$ over the bosonic twistor space. This is a super-CY, since it has a nowhere vanishing holomorphic volume form,  $\Omega = \epsilon_{\mu\nu\rho\sigma}\epsilon_{ABCD} \, z^{\mu}dz^{\nu}dz^{\rho} dz^{\sigma} d\psi^{A}d\psi^{B}d\psi^{C}d\psi^{D}$. 
The connection $\CA$  becomes a superfield and its component expansion 
\begin{equation} 
   {\cal A} = d\bar z  \left(A + \psi^{I} \chi_{I} + \psi^{I} \psi^{J} \phi_{IJ} + \psi^{I} \psi^{J} \psi^{K} \epsilon_{IJKL} \tilde\chi^{L} + \psi^{4} G \right) 
\end{equation}
yields the ${\cal N}=4$ SYM multiplet in a helicity basis. The Chern-Simons action reproduces the action for self-dual YM fields, and the complete action is recovered by including the contributions of D1 branes wrapped on rational curves in $\CZ$. The simplest case is when the curves have degree 1; they are specified by equations 
\begin{equation}
   \omega_{\dot \alpha} =  x_{\alpha \dot \alpha} \lambda^{\alpha} \,,  \quad 
   \psi^{I} =  \theta^{I}_{\alpha} \lambda^{\alpha} \,.
\end{equation}

These equations admit a dual interpretation: a point $(x,\theta)$ in superspace defines a fibre of the twistor space, and viceversa a point $(\lambda, \omega) \in \CZ$ defines a self-dual plane in (complexified) Minkowski space\footnote{We ignore here all the issues concerning the signature of spacetime.}.  
It may be surprising that twistors, in principle well apt to describe self-dual field configurations, may turn out to be the most useful way of describing the complete perturbative sector of Yang-Mills. In fact, the breaking of parity invariance is somewhat unsatisfactory, and can be avoided by using the ambitwistor construction \cite{ambitwistor}. 

The idea underlying the ambitwistor construction is the fact that the classical equations of motion of 
${\cal N}=4$ SYM follow from the condition of integrability of gauge fields on
supersymmetric null lines. A null line is the intersection of a self-dual and an anti-self-dual plane, and so is defined by two points in two copies of ${\mCP}^{3}$ with a condition to ensure that the intersection of the corresponding planes is not empty. The suitably supersymmetrized space turns out to be a quadric in
${\mCP}^{3|3}\times{\widetilde{\mCP}}^{3|3}$:
\begin{equation}\label{quadric}
\CN = \{\omega_{\dot \alpha} \tilde \omega^{\dot \alpha} - \lambda^{\alpha} \tilde\lambda_{\alpha} + \psi^{A} \tilde\psi_{A} =0 \, \}\, .
\end{equation}

This is again a super-CY, and one could envision finding a topological string that lives on this space and reproduces the full YM amplitudes. This seems to be a very non-trivial problem (see attempts at its solution in \cite{Movshev, Mason}). 

As it has been argued in \cite{Neitzke} 
the S-duality of physical string theory descends to the topological sector. This implies the existence of an A-model realization of  ${\cal N}=4$ Yang-Mills, where the D1-instantons would be replaced by worldsheet instantons (and the D5-branes by NS5-branes). Given such an A-model description, one would expect the mirror B-model to realize the perturbative Yang-Mills amplitudes classically, without the help of instantons. The natural candidate for such a mirror is then the ambitwistor space.

Since the quadric $\CN$ is a hypersurface in a toric variety, it is possible to use the techniques explained in \cite{phases,HoriKT} in order to find the mirror manifold. The homogeneous coordinates of the toric variety correspond to chiral fields $\Phi_{i}$ in a $(2,2)$-susy $U(1)$ gauged linear sigma model, and mirror symmetry amounts to performing a T-duality. A degree $d$ hypersurface is described in the sigma model via a superpotential  $W= \; P \;G(\Phi_i)$, where $P$ is a field of charge $-d$. The vacuum equations set  $P=0, G=0$. We also have to use  the results of \cite{schwarz}, where it is shown that A-model observables on a hypersurface $M \subset V$ can be computed on a supermanifold that is a fermionic bundle over $V$. One gets it by simply replacing the auxiliary field $P$ with a fermionic superfield $\Psi$ of charge $d$. In trading $M$ for the bundle one is throwing away all the information about the superpotential, except its degree; this is not a problem since the parameters of $W$ are complex structure deformations to which the A-model is insensitive. 

The T-duality action on the fermions is similar to the one on bosons: the phase of $\Psi$ dualizes into a bosonic twisted chiral multiplet $Y$, such that  
$Y+\bar Y =\bar \Psi\Psi$ while the imaginary part is periodic. However one needs something more, namely two additional fermions $\eta,\chi$ to preserve the virtual (bosonic minus fermionic) dimension of the space. There must also be a superpotential $W= - q\;\Sigma(Y-\eta\chi) +e^{-Y}$, to account for all the massive excitations. 
It is now straightforward to apply these results to the hypersurface (\ref{quadric}) (see \cite{Kumar:2004dj}). Let us note that the quadric has two K\"ahler moduli $t_{1},t_{2}$ inherited from the embedding, corresponding to the sizes of the two $\mCP^{3}$. The computation yields the following partition function for the Landau-Ginzburg B-model:

\begin{align}\label{LG}
Z = & e^{t_1} \int \prod _{a=1,2}[dx_a du_a dv_a d\eta_a d\chi_a]
\prod_a \delta (u_a v_a + \eta_a 
\chi_a - x_a + 1) \, \delta (e^{t_1} (e^{-t_2} -1) x_1 x_2 -1) \,.
\end{align}

The $\delta$-functions inside the integral contain the information on
the geometry of the mirror manifold. The first thing to note is that
it has dimension $(3|4)$ as expected. In the limit $t_{2}\rightarrow 0$ the partition function (\ref{LG}) can be 
interpreted as an integral in an affine patch of $\mCP^{3|4}$, thus confirming the conjecture that the latter and the quadric are mirror partners. However for generic $t_{2}$ the equations we find describe some deformation of the space. Strictly speaking $\mCP^{3}$ is a rigid manifold and does not have complex deformations, and the same probably holds for its supersymmetric version. It is likely that one can make sense of the deformations only in an affine patch. It would be interesting to understand better these deformations, as they could give more insight into the nature of the corresponding A-model deformations.

It is worthwhile to notice in this respect that the above derivation of the mirror manifold is somewhat formal. One would like to substantiate the conjecture with the computation of observables on both sides. This would require first of all a proper definition of the observables, which has been problematic especially on the A-model side. The S-duality conjecture described above suggests that there should also be a B-model description in ambitwistor space, but, as already observed, at present this has not yet been found.

\section{A spacetime quantum foam}\label{sec:foam}

While most works on the twistor string have focused on the open string sector, it is also interesting to consider the  closed string sector, corresponding to conformal supergravity \cite{Berkovits:2004jj}. Here the correspondence rests upon a generalization of Penrose's construction that associates a complex 3-fold $\CZ$ to any 4-dimensional {\it conformally self-dual} (\ie with a self-dual Weyl tensor  $C_{abcd}$ ) Riemannian manifold $\CM$ \cite{AtiyahWI}. Specializing to the case $\CM=S^{4}$ one recovers the twistor space $\mCP^{3}$ we have used so far. One consequence is that deformations of the complex structure of $\CZ$ correspond to perturbations of the metric of $\CM$ that preserve self-duality. We will argue that the effect of adding D1 branes in $\CZ$ is equivalent to considering blow-ups of the 4-dim space $\CM$, and that this may lead to a quantum foam description of (a sector of) quantum gravity on $\CM$. 

Manifolds with self-dual Weyl tensor can be thought of as gravitational instantons, as they minimize the action $\int {\rm d} vol\, C_{abcd} C^{abcd} $ in a topological sector given by the value of the Hirzebruch signature $\tau=b_2^{+} - b_2^{-} $. In the B-model realization, D1 branes act as sources for the holomorphic form of the CY, whose periods parametrize the complex moduli. Therefore adding D1 branes should be equivalent to consider complex deformations once the backreaction is taken into account \cite{Nekrasov:2004js}. In a sense this means generalizing the twistor correspondence to non-perturbative gravitational fluctuations. It is natural to conjecture that the number of branes should correspond to the Hirzebruch signature. One can increase the value of $\tau$ by taking the connected sum $\CM \# \overline{{\mCP}^{2}}$; it is known that the resulting manifold is again conformally self-dual. We have then the picture that a gas of gravitational instanton can be described as a gas of D-branes in the B-model. 

In order to understand what our goal is, it is helpful to compare to the S-dual description: the effective action for the A-model is a theory of K\"ahler gravity in 6 dimensions, but (at least on toric manifolds) the strong coupling regime is more usefully described as a ``quantum foam'' made up of successive blow-ups of the manifold at the corners of its toric base \cite{Iqbal:2003ds}. The twistor string offers the possibility of a similar result for a version of 4-dimensional gravity that is dominated by instantons in some regime. We have explored this issue in \cite{Hartnoll:2004rv}. 

Although we conjecture that the picture presented above should be quite general, we focus on a particular class of (asymptotically flat) self-dual K\"ahler manifolds, which are blow-ups of $\mC^{2}$ at a finite number of points. Their twistor spaces are explicitly known, they have been constructed by Lebrun  \cite{lebrun}, and they have the special property of being bimeromorphic to projective spaces, which makes them amenable to investigation with algebraic-geometric techniques. In the following we outline the construction of the twistor spaces and we show how this class of examples verifies our conjecture, in the sense that the data of the K\"ahler geometry are captured by the periods of the holomorphic form on $\CZ$, in turn related to the D1-brance charge. We add some details about the supersymmetric version of the twistor spaces, which is necessary, just as in the flat case, in order to have a well-defined topological string. 

We start from the 4-fold ${\mathcal{B}}$ (a projective bundle over $\mCP^{1}\times \mCP^{1}$) obtained as a quotient of 
$\mC^7$ by the following identifications
\begin{align}\label{eq:cstar}
\left[z_0,z_1\right]  \sim & \quad \lambda [z_0,z_1] \,, \nonumber \\
\left[\zeta_0 ,\zeta_1\right]  \sim & \quad \mu [\zeta_0,\zeta_1] \,, \nonumber \\
\left[x,y,t\right]  \sim & \quad \nu [\lambda^{n-1}\mu\,
  x,\lambda\mu^{n-1}\, y, t] \,.
\end{align}
The equation 
\begin{equation}\label{eq:define}
F \equiv xy -  t^2 \prod_{j=1}^n P^j = 0 \,.
\end{equation}
defines a singular hypersurface $\widetilde{Z} \subset {\mathcal{B}} $. 
The $P^j$ are $n$ polynomials of the form $P^i = a^i_{mn} \zeta_m z_n$, and their zeroes are $n$ curves 
 $\{C^i\}_{i=1}^n$ in the base. 
We assume that the curves are nondegenerate and generic, implying that
they all mutually intersect at precisely two points. Away from the
curves the manifold $ \widetilde{Z}$ is a $\mCP^1$ fibration over $\mCP^1\times\mCP^1$. 
Over the curves the
fibration degenerates to two spheres joined at a point, as illustrated in the figure. The degeneration is regular except for the points where two curves intersect. Each pair of curves intersect at
two points, so there will be a total of $n(n-1)$ singular points.
\begin{figure}[h]
\begin{center}
\epsfig{file=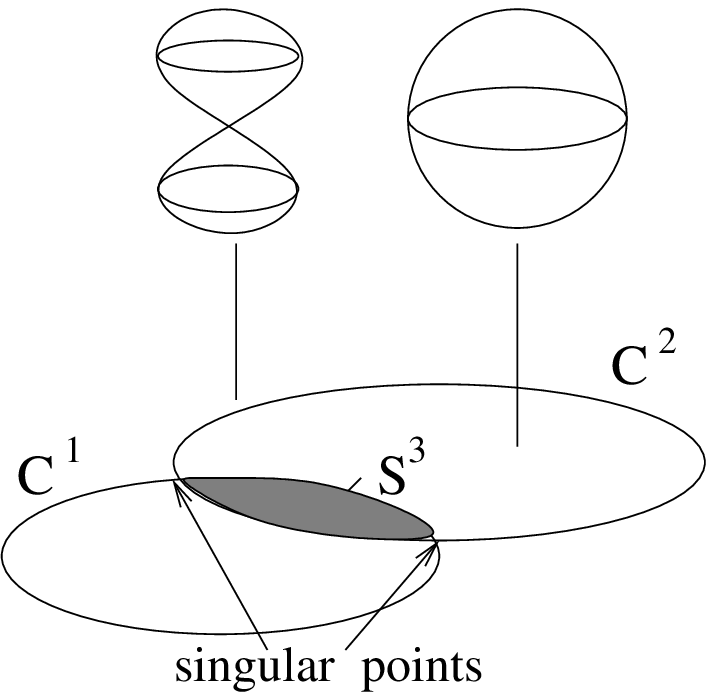,width=6cm}
\end{center}
\end{figure}

The singularities are conifold points, and they can be smoothed by taking a small resolution. The resulting manifold $\CZ$ is smooth and is a twistor space\footnote{This is true only after some additional blow-downs, we skip the details here. Note also that a deformation of the conifold singularity, as opposed to the resolution, would not give a twistor space.}, since it has a foliation by $\CP^{1}$ with normal bundle ${\cal O}(1)\oplus {\cal O}(1)$, and it has an antiholomorphic involution under which the fibres are invariant.  One can show that $\CZ$ is the twistor space of a blow-up of $\mC^{2}$ at $n$ points, and that $b_2(\CZ) =  n+1$.
The most important property of $\CZ$ for our purposes is that it contains a divisor $D$ that is a section of the twistor fibration, \ie it intersects every fibre in one point. It can be written as $\{z=c\}$ in terms of a local coordinate $z=z_{0}/z_{1}$. The involution takes it to another divisor $\bar D=\{\zeta=\bar c\}$. For twistor spaces of K\"ahler manifolds it is known that the line bundle corresponding to the divisor $D \cup \bar D$ is $K_{\CZ}^{-1/2}$. This means that if $s$ is a section of this line bundle, we can use it to write a meromorphic 3-form globally defined on $\CZ$. Explicitly we take $s=(z_{0}-c z_{1})(\zeta_{0} - \bar c \zeta_{1})$, and 
\begin{equation}\label{merom}
\Omega_{\CZ} = \frac{dz \, d\zeta \, dt}{t \, s^{2}} \, .
\end{equation}
This is in fact the Penrose transform of the K\"ahler form $\omega$ of the 4-manifold $\CM$. The Penrose transform involves a contour integral on the twistor fibres. The K\"ahler moduli of $\CM$ are given by the integrals of $\omega$ on a basis of 2-cycles $\Sigma_{i} \subset \CM$. If we take the corresponding 3-cycles $L_{i}\subset\CZ$ that include the chosen contours in the fibres, we obtain 
\begin{equation}\label{charge}
\int _{\Sigma} \omega = \int_{L} \Omega_{\CZ} \,.
\end{equation} 
This is the anticipated relation that allows us to describe the K\"ahler geometry of $\CM$ in terms of period integrals that are computed by the topological B-model on $\CZ$. We stress again that this relation is not limited to the class of examples we study, rather it holds for every self-dual K\"ahler 4-manifold. 

Two observations are in order at this point. First, we wanted to understand the complex deformations of $\CZ$ as coming from D1 brane charge, as given by 
\begin{equation}
\int \Omega = g_{s} N \,.
\end{equation}
But the number $N$ is an integer, so we might expect that the K\"ahler moduli of $\CM$ will also turn out to be quantized, exactly as it happens for the CY quantum foam of \cite{Iqbal:2003ds}. It would be interesting to make this conjecture more precise. 
Second, the 3-form we found is meromorphic. This could not be otherwise since $\CZ$ is not Calabi-Yau, but we need a holomorphic 3-form to define the B-model. We already know the solution from the flat case: it is necessary to add fermionic coordinates. There is a general recipe for producing a super-CY out of any twistor space, but for Lebrun's manifolds we can use a simpler approach. First we extend the projective bundle ${\mathcal{B}}$ adding a rank 4 fermionic vector bundle $E$, with coordinates $\eta_{i}$, and define the super-twistor space $\CZ_{S}$ by a suitable modification of the hypersurface equation. The ansatz 
\begin{equation} 
F \rightarrow {\cal F} = F + \frac{F}{t s^{2}} \eta^{1}\eta^{2}\eta^{3}\eta^{4} 
\end{equation} 
is well-defined provided the fermions are assigned the right scaling properties under (\ref{eq:cstar}). 
Then we can define a holomorphic top-form on $\CZ_{S}$ using Poincare's residue map. It is given by 
 \begin{equation}\label{superform}
\Omega_{\CZ_{S}} = \frac{\Omega_{{\mathcal B}}}{d{\mathcal F}} \,.
\end{equation}
where $\Omega_{{\mathcal B}}$ is the unique form on ${\mathcal B}$ with a pole along ${\mathcal F}=0$. 
The form (\ref{superform}) is well-defined and holomorphic, and it reduces to (\ref{merom}) if the fermions are integrated out.  This construction involved some arbitrary choices and it would be interesting to understand to what extent these choices are fixed. Note that a similar holomorphic superform has been suggested in \cite{Mason} for the ambitwistor space. 

Finally, we observe that an interesting question is whether the mirror symmetry between twistor and ambitwistor space extends beyond the flat case. Obviously it is necessary to find appropriate supersymmetric generalizations so that one has Calabi-Yaus on both sides, and for the ambitwistor space, at least to the author's knowledge, the general prescription is not known. 

I would like to thank the organizers of the workshop ``Supersymmetries and Quantum Symmetries 05'', and S. Hartnoll for commenting on the manuscript.

\end{document}